\documentclass[english,pre,a4paper,twocolumn,superscriptaddress,showpacs,amsmath,amssymb]{revtex4}
\usepackage{graphicx,color,natbib,soul}

\newcommand{\dd}{\mathrm{d}}

\newcommand{\mean}[1]{\langle #1 \rangle}
\newcommand{\Int}[1]{\int\dd #1\;}
\newcommand{\IInt}[3]{\int_{#2}^{#3}\dd #1\;}

\renewcommand{\vec}[1]{\mathbf #1}

\newcommand{\al}{\alpha}

\newcommand{\vhi}{\varphi}

\newcommand{\nois}{\boldsymbol\xi}
\newcommand{\x}{\vec r}
\newcommand{\X}{\vec R}
\newcommand{\tx}{\tau_\text{r}}
\newcommand{\Da}{D_\text{a}}
\newcommand{\Tef}{T_\text{eff}}

\newcommand{\ra}{\rightarrow}
\newcommand{\mm}{\mean{e_x}_L}
\newcommand{\mv}{\mean{e_x^2}_L}
\newcommand{\Fs}{F^\text{s}}  
\newcommand{\Fsv}{{\vec F}^\text{s}}  

\newcommand{\pwall}{p_\text{wall}}
\newcommand{\ploc}{p_\text{loc}}
\newcommand{\pvir}{p_\text{vir}}

\newcommand{\lp}{\ell_\text{p}}
\newcommand{\rhobar}{{\bar\rho}}
\newcommand{\nb}{n_\text{b}}
\newcommand{\nw}{n_\text{w}}

\begin{document}

\title{Ideal bulk pressure of active Brownian particles}

\author{Thomas Speck}
\affiliation{Institut f\"ur Physik, Johannes Gutenberg-Universit\"at Mainz,
  Staudingerweg 7-9, 55128 Mainz, Germany}
\author{Robert L. Jack}
\affiliation{Department of Physics, University of Bath, Bath, BA2 7AY, United
  Kingdom}

\begin{abstract}
  The extent to which active matter might be described by effective equilibrium concepts like temperature and pressure is currently being discussed intensely. Here we study the simplest model, an ideal gas of non-interacting active Brownian particles. While the mechanical pressure exerted onto confining walls has been linked to correlations between particles' positions and their orientations, we show that these correlations are entirely controlled by boundary effects. We also consider a definition of local pressure, which describes interparticle forces in terms of momentum exchange between different regions of the system. We present three pieces of analytical evidence which indicate that such a local pressure exists, and we show that its bulk value differs from the mechanical pressure exerted on the walls of the system. We attribute this difference to the fact that the local pressure in the bulk does not depend on boundary effects, contrary to the mechanical pressure. We carefully examine these boundary effects using a channel geometry, and we show a virial formula for the pressure correctly predicts the mechanical pressure even in finite channels. However, this result no longer holds in more complex geometries, as exemplified for a channel that includes circular obstacles.
\end{abstract}


\maketitle


\section{Introduction}

Thermal equilibrium is quite special, not the least because the same quantity--the free energy--determines the probabilities of fluctuations and the work for reversible changes~\cite{chandler}. As one consequence, in thermal equilibrium, the \emph{mechanical} pressure exerted by a fluid on its confining walls is equal to a derivative of the free energy. Moreover, pressure is an intensive quantity and is constant throughout a large system. These properties are often taken for granted and have profoundly shaped our physical intuition.

This intuition is challenged in non-equilibrium systems, where a mechanical pressure may still be measured from the forces exerted on confining walls, but is no longer calculable from a free energy. A fundamental question is whether the pressure in such a system can be related to its bulk properties, through an equation of state. This question has received considerable attention recently for active Brownian particles (ABPs)~\cite{taka14,taka14a,yan15,solo15,solo15a,wink15,bial15,taka15,taka15a,marc15,joye16}. ABPs are driven out of equilibrium due to directed motion at a constant effective force, the direction of which undergoes rotational diffusion. This model is conceptually simple, but still captures essential properties of active matter. In particular, interacting ABPs show a motility-induced phase transition strongly resembling passive liquid-gas separation but caused dynamically~\cite{yaou12,pala13,butt13,sten13,sten14,bial14,cate15,spec15}. Due to their persistence of motion, ABPs accumulate at walls~\cite{lee13a,elge13,fily15} and exert a force, \emph{viz.}, a pressure, the properties of which have been studied for several geometries~\cite{fily14a,mall14,yang14,spell15,smal15,wyso15,niko16}. {The sedimentation profile of ABPs~\cite{pala10,encu11} also enables the discussion of pressure in experiments~\cite{gino15}. Several works have conjectured that an \emph{equation of state} exists for ABPs, relating the mechanical pressure (in infinite systems) to bulk properties~\cite{taka14,taka15,solo15}. Recently, Falasco \emph{et al.} have derived an equation of state that explicitly includes the dissipation~\cite{fala16}.}

The main purpose of this article is to highlight the distinction between the mechanical pressure $\pwall$ and the \emph{local} pressure $\ploc$, which is calculated in an open subsystem, due to forces exerted by its surroundings. Both the mechanical pressure and the local pressure can also be expressed in terms of a \emph{virial} pressure. Our central result is that for ideal ABPs, the local pressure is still given by the equilibrium formula $\ploc(\x)=\rho(\x)T$ (with local density $\rho$ and Boltzmann's constant set to unity), independent of the driving. This might seem surprising because the mechanical pressure clearly depends on the active forces.   

We give several arguments why the local pressure is nevertheless a useful concept giving deeper insights into the nature of active matter. In particular, it relates to the existence of an equation of state for ABPs, for which we consider two formulations. The weak formulation states that the mechanical pressure can be predicted from the measurement of some properties (to be specified) far away from any confining walls onto which the pressure is exerted. This interpretation is followed, \emph{e.g.}, by Solon~\emph{et al.} in Ref.~\citenum{solo15}. {In a stronger formulation we further demand that the mechanical pressure is a local observable, and thus is related to the local pressure. Only in this case does one recover the properties commonly associated with the thermodynamic pressure, as becomes evident from the fact that the equation of state does not predict the observed phase boundaries~\cite{solo15}, and the existence of a negative interfacial tension~\cite{bial15}. This distinction between mechanical pressure and local pressure is also of practical importance, since the equivalence of local and mechanical pressure} underpins the use of equilibrium statistical mechanics and numerical simulations to predict the mechanical pressure from atomistic simulations employing periodic boundary conditions, which is indeed one of the first applications of modern computing machines~\cite{metr53}.

This article is organized as follows. In Sec.~\ref{sec:model} we introduce the model of non-interacting ABPs and derive the virial pressure in a large closed system following previous results. In Sec.~\ref{sec:local} we present three analytical pieces of evidence for a local pressure based on the Smoluchowski equation with explicit (although infinitely separated) walls. We then consider finite wall separations in Sec.~\ref{sec:finite} showing that mechanical and virial pressure still coincide if treating the boundary conditions correctly. In Sec.~\ref{sec:disc} we discuss the implications of our results, before concluding.


\section{Model and virial pressure}
\label{sec:model}

\subsection{Active ideal gas}

We consider an ideal (non-interacting) system of $N$ active Brownian particles at temperature $T$ in two dimensions. Each particle has a position $\x$ and an orientation $\vec e$. Particle $i$ moves according to the Langevin equation $\dot\x_i=\mu_0\vec F_i$ where $\mu_0$ is a (bare) mobility, and the total force on particle $i$ is
\begin{equation}
  \label{eq:force}
  \vec F_i = \Fsv_i + (v_0/\mu_0) \vec e_i +\nois_i,
\end{equation}
in which the forces $\Fsv_i$ come from confining walls (which may be hard
or ``soft''), $v_0$ is the speed of the active motion, and $\nois_i$ is a
thermal noise with correlations
\begin{equation}
  \label{eq:2}
  \langle \xi_i^\alpha(t) \xi_j^\beta(t') \rangle = (2T/\mu_0) \delta_{ij}\delta^{\alpha\beta} \delta(t-t'),
\end{equation}
where $\alpha,\beta$ label vector components. Throughout, we set Boltzmann's constant to unity. We consider a two-dimensional box of size $L\times L_y$. For results that are also valid in three dimensions we use $d=2$ as placeholder for the dimensionality.

\subsection{Virial pressure for closed systems}

A standard route to calculate the pressure is the virial
\begin{equation}
  \label{eq:vir}
  W \equiv \sum_{i=1}^N\mean{\vec F_i\cdot\x_i} = \sum_i \left[ 
    \mean{\Fsv_i\cdot\x_i} + (v_0/\mu_0)\mean{\vec e_i \cdot \x_i} \right].
\end{equation}
We use the It\^{o} convention so $\mean{\nois_i\cdot\x_i}=0$. Additionally,
$\partial_t\mean{|\x_i|^2}=0$ holds in the steady state of the system and
It\^o's formula yields $\sum_i\mean{\dot\x_i\cdot\x_i}+NT\mu_0 d=0$. Hence,
from $\vec F_i=\dot\x_i/\mu_0$ we find $W=-NTd$.

The first term on the right hand side of Eq.~(\ref{eq:vir}) arises from interactions of particles with the confining walls. The classical argument~\cite{hansen} assumes a pressure $\pwall$ that is isotropic and equal on all walls, and considers these walls one at a time. We place the origin in the lower left corner of the box and consider forces from the wall at $x=L$: the average total force from that wall $\sum_i\langle\Fsv_i \rangle_{L}$ points along the $(-x)$ direction with magnitude $\pwall L_y$. All particles that interact with the wall have $x\approx L$ so this wall contributes $-\pwall LL_y$ to the virial. The wall at $y=L_y$ gives an identical contribution. Combining these results and using the fact that particles evolve independently predicts that the wall pressure is given by the virial pressure,
\begin{equation}
  \pwall = \pvir^\text{closed} \equiv \rhobar T  + \frac{v_0\rhobar}{\mu_0
    d}\mean{\vec e_i\cdot\x_i},
  \label{equ:p-wall}
\end{equation}
where $\rhobar\equiv N/(L L_y)$ is the average density. For a diffusive ideal gas ($v_0=0$) we recover the ideal gas law $\pwall=\rhobar T$. In general, notice that Eq.~(\ref{equ:p-wall}) is a relation between the pressure \emph{at the walls} and the \emph{average} density in a \emph{closed} system.

\subsection{Freely diffusing orientations}

The derivation of Eq.~(\ref{equ:p-wall}) made no assumptions about the dynamical evolution of the orientation $\vec e$. In the simplest ABP model, we assume that this vector undergoes free rotational diffusion with correlation time $\tx$, so that
\begin{equation}
  \mean{\vec e(t)\cdot\vec e(t')} = e^{-|t-t'|/\tx}
\end{equation}
and the stochastic dynamics of the system is Markovian.

In two dimensions ($d=2$), we write the particle orientation as $\vec e=(\cos\vhi,\sin\vhi)^T$. The time evolution of the joint probability $\psi(\x,\vhi,t)$ of the position and orientation of a single particle is then governed by
\begin{equation}
  \label{eq:fp}
  \partial_t\psi = -\nabla\cdot\left[ v_0\vec e + \mu_0\vec{F}^s \right] \psi +
  \frac{1}{\tx}\partial_\vhi^2\psi + \mu_0 T \nabla^2 \psi.
\end{equation}
Since particles are independent, this equation fully describes our system: the
normalisation of $\psi$ is $\int \dd\x\int \dd\vhi\, \psi(x,\vhi)=N$. In the following it will be convenient to define the local density
\begin{equation}
  \label{eq:rho}
  \rho(\x) \equiv \IInt{\vhi}{0}{2\pi} \psi(\x,\vhi)  
\end{equation}
and the local polarization density
\begin{equation}
  \vec p(\x) \equiv \IInt{\vhi}{0}{2\pi} \vec e\psi(\x,\vhi)
\end{equation}
of the fluid.

\begin{figure*}[t]
  \centering
  \includegraphics{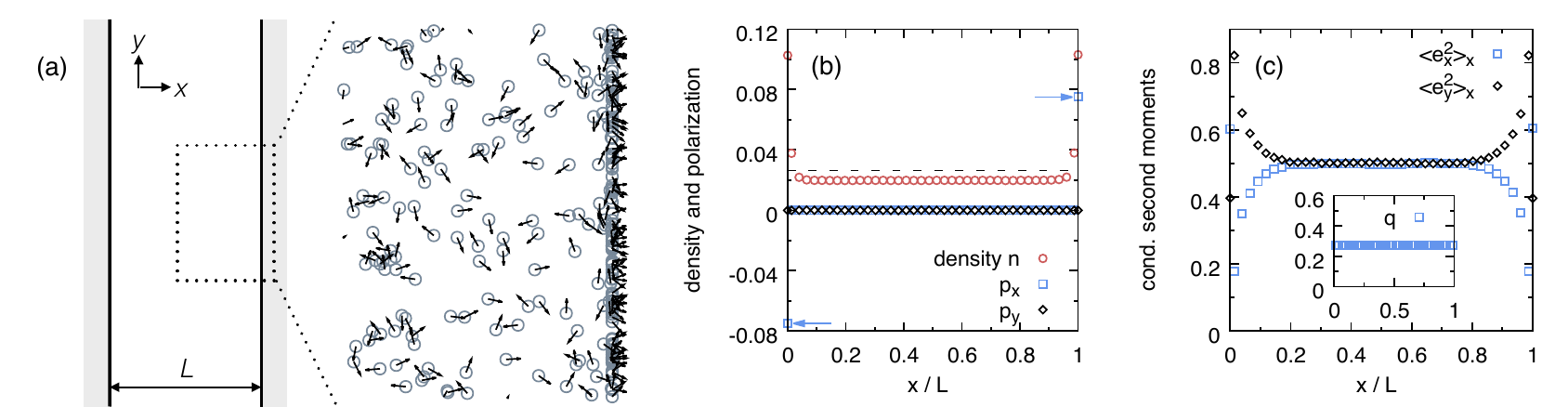}
  \caption{Channel of width $L$ with hard walls in $x$-direction and periodic boundaries in $y$-direction. (a)~Sketch of channel and snapshot of particles close to one wall for reduced inverse channel width $\nu=v_0\tx/L=0.5$, where arrows indicate the orientation. Note the adsorbed layer at the wall. (b)~Density profile and spatially resolved average orientations for $\nu=0.2$. Shown is the number of particles $n(x)\propto\rho(x)$ in uniformly spaced bins. The dashed line indicates the average density. The orientation ($p_x$ and $p_y$) is zero everywhere except exactly at the walls (arrows). (c)~Conditional second moments $\mean{e_x^2}_x$ ($\square$) and $\mean{e_y^2}_x$ ($\diamond$) for $\nu=0.5$. The inset shows that $q(x)=q$ is constant away from the walls.}
  \label{fig:walls}
\end{figure*}

\subsection{Virial pressure for a large closed system}

As a first step, from Eqs.~(\ref{equ:p-wall}) and~(\ref{eq:fp}) we calculate the mechanical pressure in a large confined system. To this end we consider the correlation function
\begin{equation}
  c(t) \equiv \langle \vec e_i(t) \cdot \vec r_i(t) \rangle = \Int{\x}\IInt{\vhi}{0}{2\pi} (\vec e\cdot\x)\psi(\x,\vhi,t).
\end{equation}
The $\x$-integral runs over the entire $x,y$ plane. The system is confined by the walls so $\psi\to0$ as $|\x|\to\infty$. The equation of motion for this correlation function is 
\begin{equation}
  \partial_t c(t) = \Int{\x}\IInt{\vhi}{0}{2\pi} (\vec
  e\cdot\x)\partial_t\psi(\x,\vhi,t).
\end{equation}
Using Eq.~(\ref{eq:fp}) and integration by parts yields
\begin{multline}
  \partial_t c(t) = \Int{\x}\IInt{\vhi}{0}{2\pi} \\ \times \psi \left[
    ( v_0\vec e + \mu_0 \vec{F}^s ) \cdot \nabla(\vec e\cdot\x)  + 
    \frac{1}{\tx}\partial_\vhi^2(\vec e\cdot\x)\right].
\end{multline}
There are no boundary terms because $\psi\to0$ at large distances, and the integrand is periodic in $\vhi$. Evaluating the derivatives yields $\partial_tc(t)=\mean{v_0 (\vec e\cdot \vec e) + \mu_0 (\vec e \cdot \Fsv)}-c(t)/\tx$. In a large system, the fraction of time that any particle spends in contact with the wall vanishes, so $\mean{\vec e\cdot\Fsv}\to0$. Hence, in steady state one has $c(t)=v_0\tx\equiv\lp$ independent of $t$. The directed motion is characterized by the persistence length $\lp$, which is the typical length over which particle orientations persist. Plugging this result into Eq.~(\ref{equ:p-wall}) yields~\cite{taka14,taka14a,yan15,solo15,solo15a,bial15,wink15,taka15,taka15a}
\begin{equation}
  \label{eq:p0}
  \pwall(\rhobar,T,v_0) = \rhobar T + p_0, \qquad p_0 \equiv
  \frac{v_0^2\tx\rhobar}{d\mu_0},
\end{equation}
which we will show below can be understood as an equation of state in the weak sense. {Two alternative formulations of this result can be given:} (i)~The active contribution $p_0=(\rhobar/d)f_0\lp$ can be rewritten as a dissipation per area, where $f_0\equiv v_0/\mu_0$ is an effective active force. {(ii)~The diffusion coefficient of non-interacting ABPs is enhanced by the active contribution $\Da=v_0^2\tx/d$~\cite{hows07}. Employing the Einstein relation, one thus finds an ideal gas law $\pwall=\bar\rho\Tef$ with effective temperature $\Tef=T+\Da/\mu_0$.}


\section{Local pressure}
\label{sec:local}

To demonstrate that the concept of a local pressure is still useful even for active particles, we abandon the closed system and consider an infinite channel of width $L$ bounded by two walls, see Fig.~\ref{fig:walls}(a). To recover the limit of a large system, we formally let $L\to\infty$. Note that all numerical data presented in the following correspond to $T=0$, while for the analytical calculations we retain a finite temperature $T$. In Fig.~\ref{fig:walls}(a) we clearly see that particles adsorb at the wall with particle orientations pointing towards the wall. The system is translationally invariant so that $\psi(x,\vhi,t)$ is independent of $y$ and Eq.~(\ref{eq:fp}) simplifies to
\begin{equation}
  \label{eq:fp:ch}
  \partial_t\psi = -\nabla_x(v_0e_x + \mu_0\Fs_x - \mu_0 T\nabla_x)\psi + 
  \frac{1}{\tx}\partial_\vhi^2\psi.
\end{equation}
Integration over $\vhi$ yields a steady-state condition that corresponds to the current being zero,
\begin{equation}
  \label{eq:j}
  v_0 p_x + \mu_0 \rho\Fs_x - \mu_0\nabla_x (\rho T) = 0,
\end{equation}
where $p_x(x)$ is the $x$-component of the polarisation $\vec p(x)$ (there should be no confusion with the pressure). We see immediately that for $T=0$ then $\vec p=0$ except at the walls, as shown numerically in Fig.~\ref{fig:walls}(b). To the extent that $\rho T$ is a local pressure, this equation corresponds to hydrostatic equilibrium ($\rho \vec f-\nabla\ploc=0$, where $\vec f$ is a one-body force per particle). Here the external forces and the internal polarisation density provide the relevant one-body forces, which are balanced by pressure gradients~\cite{yan15,thom15}. We stress that this equation does not depend on the dynamics of the orientation $\vec e$, it is valid even if torques are present. This is our first piece of evidence that $\rho T$ might serve as local pressure even for ABPs.

In addition, multiplying by $e_x$ in Eq.~(\ref{eq:fp:ch}) and then integrating
over the orientation, we have (in steady state)
\begin{equation}
  \label{eq:p}
  \nabla_x(v_0 q + \mu_0\Fs_x p_x - \mu_0 T\nabla_xp_x) + p_x/\tx = 0,
\end{equation}
where
\begin{equation}
  \label{eq:q}
  q(x) \equiv \IInt{\vhi}{0}{2\pi}e_x^2\psi(x,\vhi)
\end{equation}
is the second moment. For $T=0$ we have from Eq.~(\ref{eq:j}) that $p_x=0$ except at the walls. Hence, from Eq.~(\ref{eq:p}), $q=q(x)$ is constant everywhere except at the walls [inset to Fig.~\ref{fig:walls}(c)]. Fig.~\ref{fig:walls}(c) shows the conditional second moments $\mean{e_x^2}_x=q/\rho(x)$ and $\mean{e_y^2}_x=1-\mean{e_x^2}_x$. In the center of the channel we find $\mean{e_x^2}_x\simeq\tfrac{1}{2}$ corresponding to uniformly distributed orientations since
\begin{equation}
  \label{eq:center}
  \IInt{\vhi}{0}{2\pi} \cos^2\vhi P(\vhi) = \frac{1}{2}
\end{equation}
with $P(\vhi)=(2\pi)^{-1}$. Approaching the walls, the distribution of $e_x$ changes, due to the exchange of particles between bulk and the adsorbed layers.

Using Eq.~(\ref{eq:j}) to eliminate the non-gradient term in Eq.~(\ref{eq:p}), we obtain $\rho\Fs_x=\nabla_xG$ with
\begin{equation}
  \label{eq:G}
  G(x) \equiv \rho T+v_0\tx(p_x\Fs_x+v_0q/\mu_0-T\nabla_xp_x),
\end{equation}
which takes the role of an effective potential. This relationship enables two useful calculations. First, following Solon \emph{et al.}~\cite{solo15}, we integrate from the center of the channel ($x=L/2$) to a point outside the system ($x\to\infty$). The mechanical pressure on the right wall then reads
\begin{align}
  \label{equ:pwall-q}
  \pwall &= -\IInt{x}{L/2}{\infty} \rho \Fs_x = G(L/2) \\
  \nonumber
         &= \rho(L/2)T + \frac{v_0^2 \tx}{\mu_0}  q(L/2),
\end{align}
where we used $G\to0$ as $x\to\infty$ and $p_x(L/2)=0$ (by symmetry). This result relates the wall pressure to a local measurement of $\rho$ and $q$, which thus fulfills our weak formulation for an equation of state. Eq.~(\ref{equ:pwall-q}) still depends on the distance from the wall. For a wide channel we let $L\to\infty$ with $\rho(L/2)=\rhobar$ and $q(L/2)=\rhobar/2$ [cf. Eq.~(\ref{eq:center})], which leads to the same pressure Eq.~(\ref{eq:p0}) as in a large closed system~\cite{taka14,taka14a,yan15,solo15,solo15a,wink15}. As discussed in Ref.~\citenum{solo15a}, the derivation is valid only if the walls do not exert torques on the particles, in which case the wall pressure would still depend on properties of the boundary~\cite{joye16}. (That is, we require that $\vec e_i$ undergoes free rotational diffusion for all particles, even if they are in contact with the wall.) Our discussion highlights that this result hinges on a force density that can be written as a gradient, \emph{viz.}, the existence of a potential Eq.~(\ref{eq:G}).

However, note that Eq.~(\ref{equ:pwall-q}) is not the only possible result. Consider a hard wall for $T>0$ and integrate $\rho \Fs_x=\nabla_x G$ from $x=L/2$ to $x=L^-$ just inside the wall so that $G(L^-)-G(L/2)=0$. At the wall we have the boundary condition $v_0q/\mu_0=T\nabla_xp_x$~(see appendix~\ref{sec:bound}). With $G(L/2)=G(L^-)=\rho(L^-)T$ we thus arrive at
\begin{equation}
  \label{equ:rho-wall}
  \pwall = (\rho T)_\text{wall},
\end{equation}
which is our second piece of evidence that $\rho T$ can be interpreted as a local pressure: the same quantity that plays the part of the local pressure in Eq.~(\ref{eq:j}) also yields the mechanical pressure when evaluated at a (hard) wall. In the limit $T\to0$ the density diverges at the wall, maintaining a finite value of $(\rho T)_\text{wall}$.

\begin{figure}[b!]
  \centering
  \includegraphics[scale=.92]{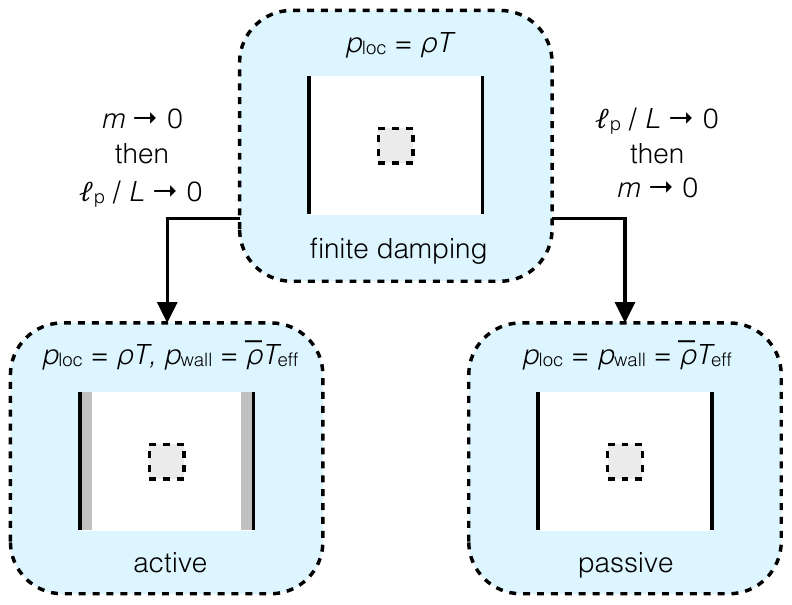}
  \caption{The order of limits matters. Starting from dynamics with finite damping and finite speed $v_0>0$, the overdamped limit $m\to0$ (with mass $m$) and the limit $\lp/L\to0$ do not commute. Taking the overdamped limit first (left) leads to ABPs in a large system, for which there is an adsorbed layer (dark stripes) and thus local and wall pressure are different. Taking the overdamped limit last (right), one obtains a passive system with uniform density, in which local and wall pressure agree (with effective temperature $\Tef$). Note that the wall pressures for both the active and passive overdamped system coincide.}
  \label{fig:limits}
\end{figure}

The third piece of evidence that $\rho(\x)T$ is the most appropriate definition of the local pressure comes from considering a Langevin system with finite friction, for the detailed calculation see appendix~\ref{sec:mom}. In this case one has a straightforward definition of the local pressure tensor in terms of momentum exchange with the environment of a localised subsystem: in the overdamped limit this reduces to the local pressure $\rho(\x)T$. This result is independent of the active forces because, while the active forces do affect the velocity, they have negligible contribution to the momentum flux in the overdamped limit. Eq.~(\ref{eq:j}) also has a natural interpretation as force balance (hydrostatic equilibrium) in this case. Note that this result for the local pressure requires that we take the overdamped limit with a fixed value of the orientational diffusion time $\tx$. If one alternatively takes $\tx\to0$ (with fixed $D_\text{a}=v_0^2\tx/d$) \emph{before} taking the overdamped limit, the system becomes microscopically reversible and one recovers simple (passive) diffusion with effective temperature $\Tef=T+D_\text{a}/\mu_0$, see Fig.~\ref{fig:limits}. While this is the same effective temperature as in the large closed system [Eq.~(\ref{eq:p0})], the active and passive systems differ qualitatively: in the passive system, there are no adsorbed layers at the walls of the system, and the (local) pressure is equal to the wall pressure $\rho\Tef$.


\section{Pressure in finite systems}
\label{sec:finite}

\subsection{Mechanical pressure}

We now explicitly calculate the mechanical pressure exerted by the active particles onto the walls for any value of the channel width $L$. The geometry of the channel affects the system through a single dimensionless control parameter, the reduced inverse channel width $\nu\equiv\lp/L$. For clarity we also take $T=0$, so particles move only in response to active forces. We recall Fig.~\ref{fig:walls}(a), which shows that particles in the channel form an adsorbed layer at the boundaries, and that the particles in the layer have orientations pointing towards the wall. The non-uniformity of the density is also shown in Fig.~\ref{fig:walls}(b). Although the system is quite simple, a full analytical solution for the density profile is already out of reach~\cite{lee13a,elge13} and we solve the equations of motion numerically.

For $T=0$ we make the ansatz
\begin{equation}
  \label{eq:psi-dirac}
  \psi(x,\vhi) = \psi_0(x,\vhi) + \frac{N\nb}{L_y}
  [\delta(x-L) P(\vhi) + \delta(x) P(-\vhi)],
\end{equation}
where $\psi_0(x,\vhi)$ is a smooth function and the Dirac delta functions account for the adsorbed layers. The fraction of particles in each adsorbed layer is $\nb$ and $P(\vhi)$ is normalised as $\Int{\vhi}P(\vhi)=1$. Fig.~\ref{fig:walls}(b) shows that the polarisation $\vec p(\x)$ is zero everywhere except for the adsorbed layers, as predicted by Eq.~(\ref{eq:j}), since $T=0$. The average orientation of a particle within the layer at $x=L$ is $\mm=\Int{\vhi}\cos\vhi P(\vhi)$. For the layer at $x=0$ then $\mean{e_x}_0=-\mean{e_x}_L$.

The pressure on the walls can be analysed in terms of the adsorbed layers.
Each particle in such a layer exerts a force $f=f_0e_x=(v_0/\mu_0)e_x$ on the wall so the pressure is
\begin{equation}
  \label{eq:p:m}
  \pwall = \mean{f} = \frac{v_0\rhobar}{\mu_0} \nb L\mm
\end{equation}
employing the distribution Eq.~(\ref{eq:psi-dirac}). Numerically, for the moments we find $\mm\simeq0.73$ and $\mv=\Int{\vhi}\cos^2\vhi P(\vhi)\simeq0.60$, independent of $\nu$.

\begin{figure}[t]
  \centering
  \includegraphics{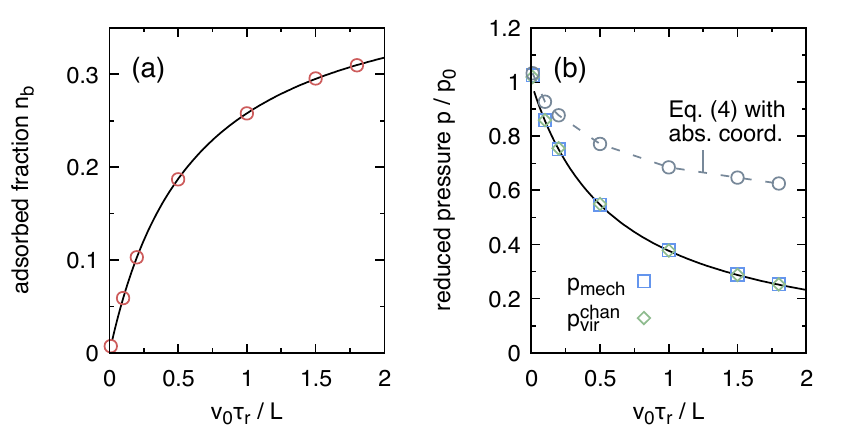}
  \caption{(a)~Fraction $\nb$ of particles adsorbed at each wall (\emph{i.e.}, their orientations point into the wall) as a function of reduced inverse channel width. (b)~Reduced pressure. Shown are the numerical results (symbols) for the mechanical pressure [Eq.~(\ref{eq:p:m})] and the virial pressure [Eq.~(\ref{eq:p:ch})] together with the analytical prediction [Eq.~(\ref{eq:p:ch:2}), line], which show excellent agreement. In contrast, evaluating the virial pressure from Eq.~(\ref{equ:p-wall}) employing absolute positions overestimates the mechanical pressure (see Sec.~\ref{subsec:per}). In all cases, the ratio $p/p_0$ approaches unity for infinite systems, $\nu\ra0$.}
  \label{fig:finite}
\end{figure}

The number of particles trapped at each wall can be determined from a two-state model, where a particle is adsorbed with rate $k_{1\ra0}\sim v_0/L$ and goes back to the bulk with rate $k_{0\ra1}\sim1/\tx$.  Balancing adsorption and desorption rates, $\nb k_{0\ra1}=(n_\infty-\nb)k_{1\ra0}$, we obtain $\nb(\nu)=n_\infty/(1+\al/\nu)$, the agreement of which with the numerical results is excellent, see Fig.~\ref{fig:finite}(a), with fitted $n_\infty\simeq0.41$ and $\al\simeq0.60$~\footnote{If a two-state description held exactly, one would have $n_\infty=0.5$, so this simple description is not perfect, but the good fit in Fig.~\ref{fig:finite}(a) indicates that it does capture the essential physics.}. Substituting these results in Eq.~(\ref{eq:p:m}), we obtain
\begin{equation}
  \label{eq:p:ch:2}
  \pwall(\nu) = \frac{p_0}{1+\nu/\al}
\end{equation}
with $p_0$ as defined in Eq.~(\ref{eq:p0}), which is recovered in the large-system limit $\nu=\lp/L\to0$ corresponding to an effectively thermalized gas with a pressure $p_0$. As expected, the pressure in a wide infinite channel is the same as that of a large closed system. While the reduction of mechanical pressure for narrow channels has been noted in numerical studies~\cite{ezhi15,yan15a}, its connection to correlations has not been discussed so far.

\subsection{Periodic boundaries and the winding number}
\label{subsec:per}

Before discussing the correlations we go back to the virial pressure and derive the analog of Eq.~(\ref{equ:p-wall}) for the channel. To this end we multiply Eq.~(\ref{eq:fp:ch}) by $x^2$. On integration by parts (twice) with respect to $x$, boundary terms vanish due to confinement of the system and in a steady state we have
\begin{equation}
  0 = 2\mu_0\mean{x\Fs_x} + 2v_0\mean{xe_x} + 2\mu_0 T.
\end{equation}
One identifies $N\mean{x\Fs_x}=-\pwall LL_y$, where $\pwall$ is the mechanical
pressure, leading to
\begin{equation}
\label{eq:pvir-chan}
  \pwall = \rhobar T + \frac{v_0\rhobar}{\mu_0}\mean{xe_x}.
\end{equation}
Due to the translational invariance the correlations $\mean{ye_y}\propto\mean{e_y}=0$ vanish, leading to a virial pressure
\begin{equation}
  \label{eq:p:ch}
  \pvir^\text{chan} \equiv \rhobar T + \frac{v_0\rhobar}{(d-1)\mu_0}\mean{\vec
    e\cdot\x},
\end{equation}
with $\pwall=\pvir^\text{chan}$ for the infinite channel. It differs from Eq.~(\ref{equ:p-wall}) for the closed system through the factor of $d-1$ on the right hand side. We have performed numerical simulations to obtain the mechanical pressure $\pwall$ (from the forces of adsorbed particles) and to evaluate Eq.~(\ref{eq:p:ch}) by calculating the correlations from all particles. The result is plotted in Fig.~\ref{fig:finite}(b) as a function of $\nu$, showing that both pressures indeed agree.

\begin{figure}[t]
  \centering
  \includegraphics{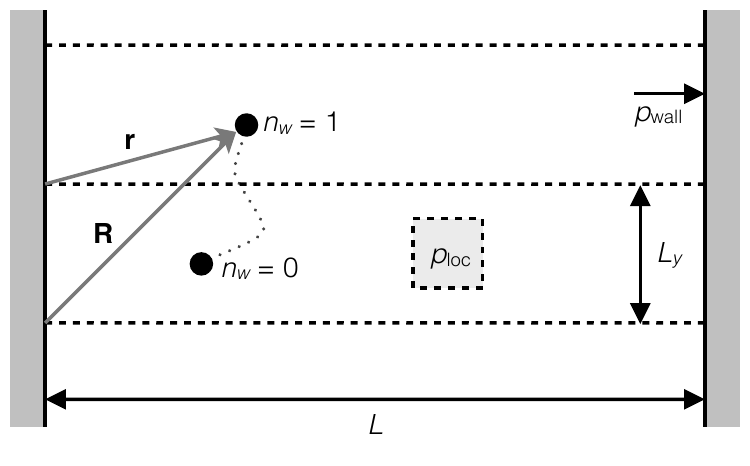}
  \caption{Channel bounded by two hard walls. We consider a box with dimensions $L\times L_y$ bounded by two hard walls and periodic boundary conditions in $y$-direction. The two pressures are indicated, the mechanical pressure $\pwall$ exerted on the walls and the local pressure $\ploc$ defined for a small subsystem away from the walls. Sketched is the trajectory of a single particle crossing the boundary of the box. Its position can thus be described by either $\x$ within the periodic box or the absolute position $\X=\nw L_y\vec e_y+\x$ with winding number $\nw$.}
  \label{fig:winding}
\end{figure}

Since the mechanical pressure is the same for the closed system and the channel, the difference between Eq.~(\ref{equ:p-wall}) and Eq.~(\ref{eq:p:ch}) means that the correlation function $\mean{\vec e\cdot\x}$ must be different in these two geometries. Some care is required in analysing this correlation function, due to its dependence on boundary conditions: this fact is intrinsically related to the dependence of the active pressure on particle ordering near the boundaries (walls) of the system. To illustrate this, note that our simulations employ a finite box with dimensions $L\times L_y$. To simulate the infinite channel, we introduce periodic boundaries in $y$-direction, see Fig.~\ref{fig:winding}. When calculating the correlations $\mean{\vec e\cdot\x}$ numerically we take the position $\x$ as the \emph{periodic} position, \emph{within} the simulation box. This is the typical situation for calculating pressures in passive simulations. It ensures that the pressure can be determined from configurations alone and thus is a local observable.

In contrast, it has previously been argued that one should evaluate $\mean{\vec e\cdot\x}$ by taking $\x\to\X$ as the \emph{absolute} position $\X=\nw L_y\vec e_y+\x$, which takes into account the crossing of periodic boundaries~\cite{taka14,wink15}, see Fig.~\ref{fig:winding}. Practically, this means that one has to introduce a winding number $\nw$ counting the number of periodic boxes a particle has traversed, analogous to the calculation of, \emph{e.g.}, the mean-square displacement. The winding number clearly is not a local observable (it depends on the history of the particle). Therefore, even if the pressure of the system can be written in terms of the correlation function $\mean{\vec e\cdot\x}$ that is evaluated in this way, this does not constitute an equation of state even in the weak sense, since the estimate for the pressure depends on the histories of all particles in the system. In Fig.~\ref{fig:finite}(b) we have plotted the pressure employing Eq.~(\ref{equ:p-wall}) with absolute positions along the channel (including winding numbers). As $\lp/L\to 0$, the correct result for a large system is recovered. However, for finite channel width $L$ employing the absolute positions overestimates the mechanical pressure.

Note also that while Eq.~(\ref{eq:p:ch}) generalises Eq.~(\ref{equ:p-wall}) to the channel geometry, there is no direct generalisation to a fully periodic system without introducting winding numbers or other history-dependent terms. This is in contrast to the equilibrium virial formula for the pressure in systems of interacting particles, where the analog of the $\mean{\vec e\cdot\x}$ term depends only on separations between pairs of particles, and is easily estimated from simulations of periodic systems.

\subsection{Active pressure is a boundary effect}

We finally stress that the active pressure is controlled entirely by boundary effects. Evaluating the correlations $\mean{\vec e\cdot\x}$ in a small volume $B$ away from the boundaries leads to
\begin{equation}
  \mean{\vec e\cdot\x}_B = \IInt{\x}{B}{} \vec p(\x)\cdot\x = 0
\end{equation}
since $\vec p=0$, cf. Fig.~\ref{fig:walls}(b). Hence, in agreement with the local pressure being independent of the active forces, ABPs do not exert a pressure by themselves but only due to their accumulation at walls caused by the directed motion. This means that the \emph{bulk pressure} (that is, the local pressure far from any boundaries) is equal to $\rhobar T$ and vanishes in the limit $T\to0$, for which all particle motion comes from active forces. While this might seem surprising, the reason is that the active forces do not contribute to the momentum flux
(see Appendix~\ref{sec:mom})

\subsection{Obstacles}

\begin{figure}[t]
  \centering
  \includegraphics{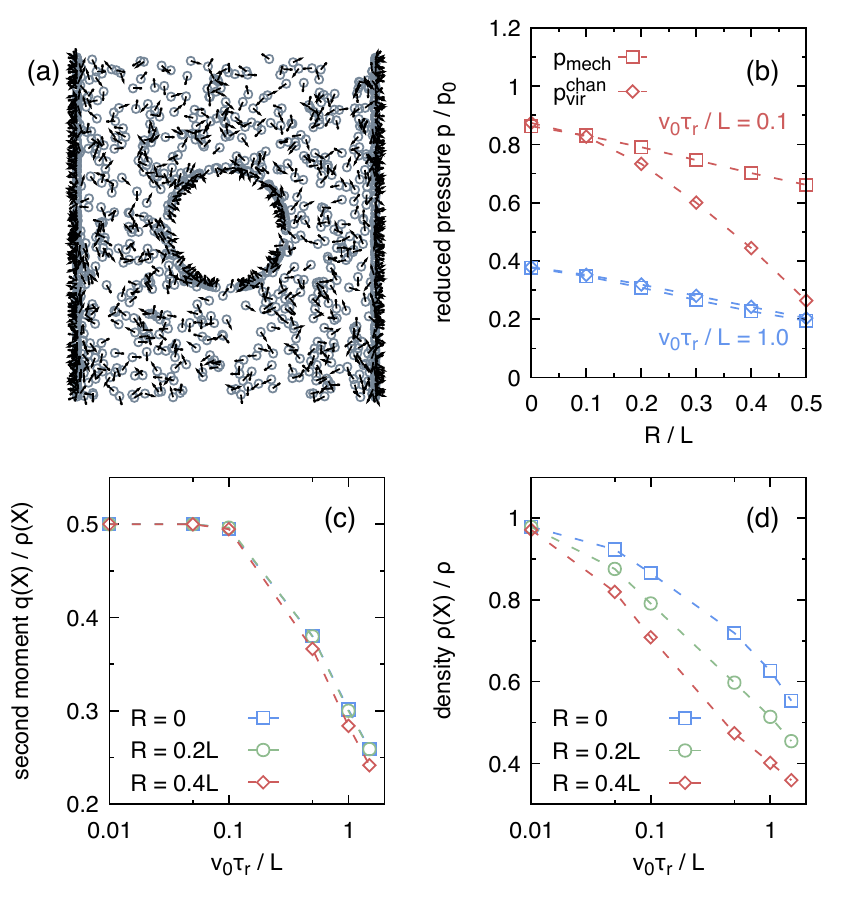}
  \caption{Channel with obstacles. (a)~Snapshot for $\nu=0.5$ and obstacle radius $R/L=0.2$. (b)~Reduced pressure for two values of $\nu$ as a function of obstacle radius $R/L$. (c)~Conditional second moment for a plane $x=X$ close to the walls. (d)~Reduced density $\rho(X)/\rhobar$.}
  \label{fig:obst}
\end{figure}

Finally, we discuss what happens when the environment in which the particles move is changed via the introduction of a second length scale. To this end we place hard circular obstacles with radius $R$ along the center of the infinite channel, see Fig.~\ref{fig:obst}(a) for a snapshot. The distance between obstacles is $L_y$. We still employ periodic boundary conditions for the $y$-direction with one obstacle per box and adjust $L_y$ so that the global density $\rhobar$ remains unchanged. Hence, we now have two reduced lengths, $\lp/L=\nu$ as before and $R/L$.

Fig.~\ref{fig:obst}(a) shows that the particles accumulate at the obstacle, as expected. In Fig.~\ref{fig:obst}(b) we plot the reduced mechanical pressure $\pwall/p_0$ exerted onto the two outer walls as a function of radius $R/L$ for two values of $\nu$. It shows that the presence of the obstacle reduces the forces on the outer walls. We have also calculated the virial pressure $\pvir^\text{chan}$ [Eq.~(\ref{eq:p:ch})], which for small $\nu$ deviates from the mechanical pressure. Hence, in the presence of internal walls the correlations are not sufficient anymore to capture the mechanical pressure.

We now generalize the approach of Sec.~\ref{sec:local}. The system is no longer translationally invariant. However, all quantities are periodic functions with respect to the $y$ coordinate. The mechanical pressure [cf. Eq.~(\ref{equ:pwall-q})] reads
\begin{equation}
  \pwall = -\frac{1}{L_y}\int_0^{L_y}\dd y\IInt{x}{X}{\infty} \rho\Fs_x,
  \label{equ:pw-obst}
\end{equation}
where the lower limit of integration is now a plane $x=X$ outside the obstacle, $(L/2+R)<X<L$. While Eq.~(\ref{eq:j}) still holds, the force density now reads $\rho\Fs_x=\nabla_xG+\nabla_yG_y$ with potential $G=G(\x)$ similar to that in Eq.~(\ref{eq:G}) and an additional potential term $G_y(\x)$. The latter term does not contribute to $\pwall$ due to periodicity of the $y$-integral in Eq.~(\ref{equ:pw-obst}). Using again that $G\to0$ as $x\to\infty$, we obtain
\begin{equation}
  \pwall = \frac{1}{L_y}\IInt{y}{0}{L_y} G(X,y).
\end{equation}
For $T=0$ we numerically find $p_x=0$, away from the walls~\footnote{In contrast to the translationally invariant channel where $p_x=0$ follows from Eq.~(\ref{eq:p}), deriving the equivalent result in the presence of the obstacle requires an assumption that there are no persistent currents in the steady state of the system. This condition is in fact satisfied within our numerics, so $p_x=0$.}. Hence the mechanical pressure reduces to
\begin{equation}
  \pwall = \frac{v_0^2\tx}{\mu_0}\bar q(X),
\end{equation}
where the second moment $\bar q(X)$ is evaluated at the plane $x=X$, and is averaged over the $y$ coordinate. Fig.~\ref{fig:obst}(c) shows the behavior of $\bar q(X)/\rho(X)$ for a plane close to the walls as a function of $\nu=\lp/L$ for different obstacle radii. The ratio approaches $\bar q(X)/\rho(X)\to\tfrac{1}{2}$. In Fig.~\ref{fig:obst}(d) the density $\rho(X)/\rhobar$ is plotted, which shows that in the chosen plane the density also approaches the bulk density. Both figures show that in the limit of a large system with fixed $R/L$ the same wall pressure Eq.~(\ref{eq:p0}) is recovered as in the closed system and the infinite channel without obstacles. However, the virial formula (\ref{eq:p:ch}) does not coincide with the mechanical pressure, due to the presence of the internal walls.


\section{Discussion}
\label{sec:disc}

Suppose we want to know the pressure of, say, liquid argon at a given temperature. We can perform a computer simulation employing the Lennard-Jones pair potential with a fixed number of particles and fixed box size. An instantaneous pressure is calculated from the particle positions and averaged over many configurations sampled at equilibrium. The larger the system the fewer configurations are needed to converge the numerical estimate for the pressure. Periodic boundaries are helpful here, in that they enable accurate estimates of bulk quantities from simulations of finite sytems, as long as the simulated system is large (in comparison with microscopic correlation lengths). Moreover, if a very large system is considered, the pressure can be measured by observing a finite subsystem within it, and the result is independent of the subsystem chosen. This is true even if the system is inhomogeneous, for example, if it contains liquid and vapor in coexistence.

In contrast, the mechanical pressure of active particles is intrinsically wound up with the layers of particles which form at the boundary of the system. This observation prevents estimation of the mechanical pressure in terms of state variables within periodic systems, even for the simplest case of non-interacting active Brownian particles (see Sec.~\ref{subsec:per}). To the extent that a local pressure can be defined (either in periodic systems or in finite subsystems far from walls), it is equal to $\rho T$, as shown here, and therefore differs from the mechanical pressure. In that strict sense, an equation of state does not exist for active Brownian particles. A weaker formulation [cf. Eq.~(\ref{equ:pwall-q})] still holds for torque-free spherical particles~\cite{solo15}, which relates the mechanical pressure to the second moment $q$ of orientations. It explicitly involves walls and only in the limit $\lp/L\to0$ of a large separation of these walls does $q$ become a true bulk quantity. The physical picture is that in this limit the single adsorption events onto walls become uncorrelated and the active forces thus contribute uncorrelated noise that can be described by an elevated effective temperature.

Since estimators involving winding numbers can reproduce the mechanical pressure for large systems, it might be argued that the requirement to track such history-dependent quantities is an essential price to pay for estimating pressure out of equilibrium. We note however that such an approach does not capture the properties of finite systems (Fig.~\ref{fig:finite}). Moreover, if systems are inhomogeneous (for example due to motility-induced phase separation), the properties of numerical estimators involving winding numbers can be extremely poor, since convergence of averaged values requires that typical particles spend significant time in both dense and sparse phases, during the simulated trajectories~\cite{bial15}. This fact again emphasises that the winding number is not a local observable, and that approaches based on such observables cannot be used to characterise the local pressure or stresses in the fluid. This last observation is important since a practical motivation for defining the pressure is the derivation of hydrodynamic equations (in the spirit of Navier-Stokes), in which fluid flow on macroscopic scales depends on pressure gradients, as well as other macroscopic properties such as viscosity. A local definition of pressure is essential for such an approach -- this motivates our interpretation of equations such as (\ref{eq:p}) in terms of hydrostatic equilibrium (local force balance).


\section{Conclusions}

In this article, we have argued that the local pressure of ideal active Brownian particles should be identified as $\rho(\vec r) T$. This differs from the mechanical pressure exerted on walls of the system, because the walls induce changes in local density via the formation of an adsorbed layer of particles. For hard walls, the mechanical pressure is given by $(\rho T)_\text{wall}$, which lends extra support to our interpretation. The extension of these results to ABPs that interact by a two-body potential is straightforward and will be discussed in a future publication. Our results are an important step towards a better theoretical understand of pressure and forces on immersed bodies~\cite{hard14,ray14,yan15a,ran15}, which will be crucial for future application of active particles in, \emph{e.g.}, self-assembly. It would also be useful to connect this work with recently proposed mappings between ABPs and effective equilibrium systems~\cite{fara15,magg15,marc16}, in which the pressure necessarily is local.

Two questions that remain outstanding are (i)~whether (and how) these results can be related to non-equilbrium analogs of the ``thermodynamic'' pressure defined in terms of large deviations of the density (on the hydrodynamic scale); and (ii)~can the arguments presented here be extended to systems where the orientational evolution of the particles is no longer independent of the particle positions?  Examples of this latter situation include the presence of external torques (\emph{e.g.} favoring alignment at walls), or torques arising from interactions between particles. Finally, it is not clear to what extent we can expect a general hydrodynamic description of active fluids in terms of local pressure, density, and velocity, but we feel that identifying cases in which such a description is possible is a key goal in this area~\cite{witt14,tiri15}. An appropriate definition of the local pressure is obviously vital if this is to be achieved.


\acknowledgments

We thank Hartmut L\"owen, Roland Winkler, Grzegorz Szamel, and Julien Tailleur for helpful discussions and comments. TS gratefully acknowledges financial support by the DFG within priority program SPP 1726 (grant number SP 1382/3-1).


\appendix


\section{Boundary condition at a hard wall}
\label{sec:bound}

We derive the boundary condition of the active fluid at a hard wall by taking the limit of a ``soft'' wall, for which $\Fsv$ is a finite $x$-dependent force. We consider a single wall at $x=0$, so that the force $\Fsv$ lies in the positive $x$-direction for $x<0$, with $\Fsv=0$ for $x>0$. The wall confines the system so $\psi\to0$ as $x\to-\infty$.

Take $X>0$ and consider the time derivative 
\begin{equation}
  \int_{-\infty}^{X} \mathrm{d}x \int \mathrm{d}y \,\partial_t \psi = 0
  \label{equ:rhoX}
\end{equation}
of the number of particles with orientation $\vhi$ and $x<X$, which vanishes in the steady state. Also, translational invariance along the wall means that $\psi=\psi(x,\vhi)$, so the $y$ integral in (\ref{equ:rhoX}) corresponds simply to multiplication by a constant. Using Eq.~(\ref{eq:fp:ch}) and performing the integration over $x$, we have
\begin{multline}
  \label{equ:int-over-wall}
  0 = - v_0 e_x \psi(X,\phi) + \mu_0 T \nabla_x\psi(X,\phi) + \\
  \frac{1}{\tx} \int_{-\infty}^{X} \mathrm{d}x \, \partial_\phi^2 \psi(x,\phi),
\end{multline}
where we used $\Fsv(X)=0$. Now take the hard wall limit, in which case $\psi=0$ for all $x<0$. Then take $X\to0$ (from above). For $T>0$ then $\psi$ is a smooth function so the integrand in the last term is finite in the limit, and the support of the integral goes to zero. Hence, for all $\vhi$
\begin{equation}
  v_0 e_x \psi(0,\vhi) = \mu_0 T (\nabla_x\psi)_{x=0},
\end{equation}
which is the boundary condition at the hard wall. Multiplying by $e_x$ and integrating over $\vhi$ yields the boundary condition $v_0q=\mu_0T\nabla_x p_x$.


\section{Finite damping}
\label{sec:mom}

\subsection{Momentum transfer}

We consider particles with finite mass $m$ that move with velocity $\vec v_i = \dot{\vec r}_i$, with
\begin{equation}
  m\dot{\vec v}_i = \vec F_i - (\vec v_i/\mu_0)
\end{equation}
and $\vec F_i$ given by Eq.~(\ref{eq:force}). We identify $(1/\mu_0)$ as a friction (damping) coefficient. If we take $m\to0$ then we recover the overdamped dynamics considered in the main text.

To define a local pressure, we consider a subsystem $B$ of a large system with its center at $\x_0$, cf. Fig.~\ref{fig:winding}. There are no walls inside the subsystem and particles can move into and out of the subsystem. The total momentum inside $B$ can change either due to ``direct'' forces from outside the subsystem, or due to particles carried into
(or out of) the subsystem by particles crossing the boundary (\emph{i.e.}, momentum flux). The direct contribution to the pressure is
\begin{equation}
  \label{equ:pext-def}
  p_\text{ext} = \frac{N}{A_{\rm b}} \mean{-\vec{F}^{\rm ext}_i \cdot \vec{n}_{\rm b}},
\end{equation}
where $\vec{F}^{\rm ext}_i$ is the force exerted on particle $i$ by particles that are external to subsystem $B$, $\vec{n}_{\rm b}$ is normal to the boundary of $B$, and $A_{\rm b}$ is the length (area) of the boundary being considered.

The other contribution to the pressure is the rate of momentum transfer into the system from the outside, minus the rate of momentum transfer out of the system from the inside. These two terms are equal in magnitude and together they sum to
\begin{equation}
  p_\text{mom} = m\mean{\rho(v^\perp)^2}_\text{bndy},
\end{equation}
where $v^\perp\equiv\vec{n}_{\rm b}\cdot \vec v$ is the velocity perpendicular to the boundary and the average is taken in a small region located at the boundary. This motivates the definition of a (local) momentum flux tensor $\pi^{\alpha\beta}=m\mean{\rho v^\alpha v^\beta}$, which can be evaluated locally. If the system is isotropic then one has $\pi^{\alpha\beta}=p_{\rm mom}\delta^{\alpha\beta}$ with $p_{\rm mom}=m\mean{\rho v^2}/d$.

The pressure (defined in terms of momentum exchange) for subsystem $B$ is $\ploc=p_\text{mom}+p_\text{ext}$. For ideal (non-interacting) systems as defined here, we consider a subsystem far from any walls, in which case $p_\text{ext}\to0$ and $p_\text{mom}$ can be interpreted as the local pressure. Taking the overdamped limit $m\to0$ at fixed $\tx$, we find (see below) 
\begin{equation}
m\langle \rho v^\alpha v^\beta\rangle \to \rho T \delta^{\alpha\beta} ,
\label{equ:equipart}
\end{equation} 
which is the same result as one obtains from (classical) equipartition of energy at equilibrium. Hence the local pressure is $\ploc(\x)=p_{\rm mom}(\x) = \rho(\x)T$.

On the other hand, taking the limit of small $\tx$ (at fixed $\Da=v_0^2\tx/d$), we show in the following that all non-equilibrium aspects of the problem disappear: the active swim force behaves in the same way as an extra thermal noise the particle, and one recovers equilibrium behaviour at a higher temperature $\Tef=T+\Da/\mu_0$.

\subsection{Position and velocity correlations}

\newcommand{\vv}{\vec v}
\newcommand{\rr}{\vec r}
\newcommand{\ee}{\vec e}
\newcommand{\WW}{\mathbb{W}}
\newcommand{\ppv}{{\vec p}_v}
\newcommand{\ppvh}{\hat{{\vec P}}_v}
\newcommand{\ppeh}{\hat{{\vec P}}_e}

To derive (\ref{equ:equipart}), and investigate the behaviour of correlations between particles' velocities and their orientations, we consider the equations of motion for a single active particle far from any walls. (To achieve this in practice, consider a periodic system in two dimensions, without any external forces acting. In a large closed system, we can assume that particles spend negligible tine in contact with the walls, in which case the following results still apply.)  

The Fokker-Planck equation that generalises (\ref{eq:fp}) is then
\begin{multline}
\partial_t \Psi = -\vv\cdot\nabla_{\rr}\Psi + \frac{1}{m\mu_0}\nabla_{\vv}\cdot\left[ \vv - v_0 \ee + \frac{T}{m} \nabla_{\vv} \right] \Psi \\ +
(1/\tx) \partial_\vhi^2 \Psi
\label{equ:fpv}
\end{multline}
where $\Psi = \Psi(\rr,\vv,\vhi,t)$ is the distribution function for the particle's position, velocity and orientation, 
and the vector $\ee$ has elements $(\cos\vhi,\sin\vhi)$. The gradients $\nabla_\rr,\nabla_\vv$ act on the position and velocity co-ordinates, respectively. It is convenent to write $\partial_t \Psi = \WW \Psi$ where $\WW$ is a second order differential operator that may be read off from (\ref{equ:fpv}), and we have (formally) $\Psi(t) = {\rm e}^{\WW t}\Psi(0)$.
If we also assume that $\Psi$ is independent of position $\rr$ then this property is maintained by the dynamical evolution and we can consider $\Psi=\Psi(\vv,\ee,t)$. 

As in the Heisenberg picture of quantum mechanics, we define time-dependent operators for the velocity and 
orientation as $\hat{\vv}(t)={\rm e}^{-\WW t} \hat\vv {\rm e}^{\WW t}$ and $\hat{\ee}(t)={\rm e}^{-\WW t} \hat\ee {\rm e}^{\WW t}$.
It is also convenient to define a momentum operator conjugate to the velocity $i{\ppvh} = \nabla_{\vv}$ and its time-dependent 
analog ${\ppvh}(t)={\rm e}^{-\WW t}\ppvh {\rm e}^{\WW t}$. Also $iP_\phi = \partial_\vhi$ with a similar time-dependent analog. Then one has (in terms of operators)
\begin{align}
\partial_t \hat{\vv}(t) & = [-\hat\vv(t) + v_0 \hat\ee(t)]/(m\mu_0) - 2i T \ppvh(t)/(m^2\mu_0)
\nonumber \\
\partial_t \hat{\ee}(t) & = -\hat\ee(t)/\tx + 2i\hat{p}_\vhi(t) \hat{\ee}^\perp(t)/\tx
\label{equ:Psi}
\end{align}
where $\ee^\perp$ is the vector $(\sin\vhi,-\cos\vhi)$ which is perpendicular to $\ee$.  Since we have assumed that $\Psi$ is independent of $\rr$, we also have $\partial_t \ppvh(t) = \ppvh(t)/(m\mu_0)$.  Expectation values in the steady state of the model are given by $\langle \hat A \rangle = \int\mathrm{d}\vec v\,\mathrm{d}\vec e\,\mathrm{d}\vec r\, \hat A \Psi_{\rm ss}$ where $\Psi_{\rm ss}$ is the steady-state solution of (\ref{equ:Psi}).

With this machinery set up, standard methods allow us to write down and then solve equations of motion for correlation functions: for example, considering vector components of the velocity, one has
\begin{multline}
\partial_t \langle v^\alpha_t v^\beta_{t'} \rangle 
 = [-\langle v_t^\alpha v_{t'}^\beta \rangle + v_0 \langle e_t^\alpha v_{t'}^\beta \rangle]/(m\mu_0)
  \\ - 2i T \langle P_{v,t}^\alpha v_{t'}^\beta \rangle /(m^2\mu_0)
  \label{equ:vv}
\end{multline}
By considering a similar expression for $\partial_{t'} \langle v_t^\alpha  P_{v,t'}^\beta \rangle$ one obtains
\begin{equation}
i\langle v_t^\alpha P_{v,t'}^\beta \rangle = -\delta^{\alpha\beta} \Theta(t-t') {\rm e}^{-(t-t')/(m\mu_0)}
\label{equ:vp}
\end{equation}
where $\Theta(t-t')$ is the Heaviside (step) function, and we used $\langle e_t^\alpha P_{v,t'}^\beta\rangle = 0$ and $\langle P_{v,t}^\alpha P_{v,t'}^\beta\rangle=0$, which both follow from the equation of motion for $\ppvh(t)$. 

Considering next $\partial_t \langle e_t^\alpha v_{t'}^\beta \rangle$, one has that for $t>t'$:
\begin{equation}
\langle e_t^\alpha v_{t'}^\beta \rangle =\langle e_{t'}^\alpha v_{t'}^\beta \rangle {\rm e}^{-(t-t')/\tx}
\end{equation}
To fix the prefactor, we use time translation invariance of the steady state, so that $(\partial_t + \partial_{t'}) \langle e_t^\alpha v_{t'}^\beta \rangle = 0$. Evaluating this expression as $t\to t'$, one obtains the value of $\langle e_{t'}^\alpha v_{t'}^\beta \rangle$ and hence
\begin{equation}
\langle e_t^\alpha v_{t'}^\beta \rangle = \delta^{\alpha\beta} \frac{v_0\tx }{d(\tx+ m\mu_0)} {\rm e}^{-(t-t')/\tx}, \qquad t>t'
\label{equ:ev+}
\end{equation}
where we used $\langle e^\alpha_t e^\beta_t \rangle = \delta^{\alpha\beta}/d$, since $\vec e$ is a randomly distributed unit vector.
Note that in the overdamped limit $m\to0$ then $\langle \vec e_t \cdot \vec v_{t}\rangle \to v_0$, consistent with the known behaviour of ABPs.

Finally, returning to (\ref{equ:vv}), assuming $t>t'$ and using (\ref{equ:vp},\ref{equ:ev+}), we obtain a first-order differential equation for $\langle v^\alpha_t v^\beta_{t'} \rangle$ that can be solved using an integrating factor, yielding
\begin{multline}
\langle v^\alpha_t v^\beta_{t'}\rangle = C^{\alpha\beta}{\rm e}^{-(t-t')/(m\mu_0)} 
  \\ +  \delta^{\alpha\beta}\frac{v_0^2\tx^2}{d[\tau^2- (m\mu_0)^2]}{\rm e}^{-(t-t')/\tx}
\end{multline}
where the $C^{\alpha\beta}$ are constants of integration that can be fixed using time translation invariance of the steady state, which implies $(\partial_t + \partial_{t'})\langle v^\alpha_t v^\beta_{t'}\rangle=0$. Hence,
\begin{multline}
\langle v^\alpha_t v^\beta_{t'}\rangle = \delta^{\alpha\beta}\frac{v_0^2\tx^2}{d[ \tx^2- (m\mu_0)^2]}{\rm e}^{-|t-t'|/\tx} \\
+ \delta^{\alpha\beta}\left[\frac{T}{m} - \frac{m \mu_0 v_0^2\tx}{d[ \tx^2- (m\mu_0)^2]} \right] {\rm e}^{-|t-t'|/(m\mu_0)}
\end{multline}
The structure in this correlation function arises from the coupling of the velocity to the orientation. Note that there is no divergence as $\tx \to m\mu_0$ because the singular contributions from the two terms cancel each other. For active Brownian particles, we require the overdamped limit ($m\to0$), in which case this expression simplifies, leading to a Dirac delta function that encapsulates the effect of the thermal noise, and a regular term that describes the persistent motion of the particle due to the active forces. Setting $t=t'$ and taking $m\to0$, one also obtains $m\langle v^\alpha_t v^\beta_{t}\rangle \to T \delta^{\alpha\beta}$ as asserted above. That is, the momentum flux tensor is isotropic and consistent with a local pressure $\rho T$.

For completeness, note that by considering $\partial_{t'} \langle e_t v_{t'} \rangle$, one finds that for $t'>t$,
\begin{multline}
\langle e_t^\alpha v_{t'}^\beta \rangle = \delta^{\alpha\beta} \frac{ v_0\tx }{d(\tx - m\mu_0)} {\rm e}^{-(t'-t')/\tx}
\\
+ \delta^{\alpha\beta} \frac{2m\mu_0 v_0\tx}{d[(m\mu_0)^2-\tau^2]} {\rm e}^{-(t'-t)/(m\mu_0)} 
\end{multline}
which complements (\ref{equ:ev+}). It is not immediately apparent from these expressions but we note that this correlation function is continuous at $t=t'$, and there is no divergence as $\tx \to m\mu_0$ because the singular contributions from the two terms cancel each other, as was the case for $\langle v_t v_{t'}\rangle$.  

\subsection{Order of limits}

Finally, it is instructive to consider the limit $\tx\to0$ (holding $\Da=v_0^2\tx/d$ constant), in which case it may be verified that the orientation $\ee$ acts as a Brownian noise that modifies the effective temperature of the system. In particular $\langle v_0 e^\alpha_t \cdot v_0 e^\beta_t \rangle \to \delta^{\alpha\beta}\Da\delta(t-t')$ means that $v_0 \vec e$ acts in the limit as a white noise, and $\langle v_{t}^\alpha \cdot v_0 e_{t'}^\beta \rangle \to \delta^{\alpha\beta} \Theta(t-t') \frac{2\Da}{m\mu_0} {\rm e}^{-(t-t')/(m\mu_0)}$, as expected for the correlation between a velocity and a noise force. But we emphasize that active Brownian particles do \emph{not} under any circumstances reduce to this passive system, because the ABP model is obtained by taking $m\to0$ before any limit of small $\tx$. These two limits do not at all commute, as may be easily seen from the local pressure, which is $\rho T$ for ABPs but is equal to $\rho\Tef$ for the effective passive case. Also, the adsorbed layer of particles at the wall is present for ABPs [with local density $\bar\rho(1+\Da/(\mu_0T))$, from Eq.~(\ref{equ:rho-wall})] but there is no such layer for passive particles.



\begin{thebibliography}{48}
\expandafter\ifx\csname natexlab\endcsname\relax\def\natexlab#1{#1}\fi
\expandafter\ifx\csname bibnamefont\endcsname\relax
  \def\bibnamefont#1{#1}\fi
\expandafter\ifx\csname bibfnamefont\endcsname\relax
  \def\bibfnamefont#1{#1}\fi
\expandafter\ifx\csname citenamefont\endcsname\relax
  \def\citenamefont#1{#1}\fi
\expandafter\ifx\csname url\endcsname\relax
  \def\url#1{\texttt{#1}}\fi
\expandafter\ifx\csname urlprefix\endcsname\relax\def\urlprefix{URL }\fi
\providecommand{\bibinfo}[2]{#2}
\providecommand{\eprint}[2][]{\url{#2}}

\bibitem[{\citenamefont{Chandler}(1987)}]{chandler}
\bibinfo{author}{\bibfnamefont{D.}~\bibnamefont{Chandler}},
  \emph{\bibinfo{title}{Introduction to Modern Statistical Mechanics}}
  (\bibinfo{publisher}{Oxford University Press}, \bibinfo{address}{Oxford},
  \bibinfo{year}{1987}).

\bibitem[{\citenamefont{Takatori et~al.}(2014)\citenamefont{Takatori, Yan, and
  Brady}}]{taka14}
\bibinfo{author}{\bibfnamefont{S.~C.} \bibnamefont{Takatori}},
  \bibinfo{author}{\bibfnamefont{W.}~\bibnamefont{Yan}}, \bibnamefont{and}
  \bibinfo{author}{\bibfnamefont{J.~F.} \bibnamefont{Brady}},
  \bibinfo{journal}{Phys. Rev. Lett.} \textbf{\bibinfo{volume}{113}},
  \bibinfo{pages}{028103} (\bibinfo{year}{2014}).

\bibitem[{\citenamefont{Takatori and Brady}(2014)}]{taka14a}
\bibinfo{author}{\bibfnamefont{S.~C.} \bibnamefont{Takatori}} \bibnamefont{and}
  \bibinfo{author}{\bibfnamefont{J.~F.} \bibnamefont{Brady}},
  \bibinfo{journal}{Soft Matter} \textbf{\bibinfo{volume}{10}},
  \bibinfo{pages}{9433} (\bibinfo{year}{2014}).

\bibitem[{\citenamefont{Yan and Brady}(2015{\natexlab{a}})}]{yan15}
\bibinfo{author}{\bibfnamefont{W.}~\bibnamefont{Yan}} \bibnamefont{and}
  \bibinfo{author}{\bibfnamefont{J.~F.} \bibnamefont{Brady}},
  \bibinfo{journal}{Soft Matt.} \textbf{\bibinfo{volume}{11}},
  \bibinfo{pages}{6235} (\bibinfo{year}{2015}{\natexlab{a}}).

\bibitem[{\citenamefont{Solon et~al.}(2015{\natexlab{a}})\citenamefont{Solon,
  Stenhammar, Wittkowski, Kardar, Kafri, Cates, and Tailleur}}]{solo15}
\bibinfo{author}{\bibfnamefont{A.~P.} \bibnamefont{Solon}},
  \bibinfo{author}{\bibfnamefont{J.}~\bibnamefont{Stenhammar}},
  \bibinfo{author}{\bibfnamefont{R.}~\bibnamefont{Wittkowski}},
  \bibinfo{author}{\bibfnamefont{M.}~\bibnamefont{Kardar}},
  \bibinfo{author}{\bibfnamefont{Y.}~\bibnamefont{Kafri}},
  \bibinfo{author}{\bibfnamefont{M.~E.} \bibnamefont{Cates}}, \bibnamefont{and}
  \bibinfo{author}{\bibfnamefont{J.}~\bibnamefont{Tailleur}},
  \bibinfo{journal}{Phys. Rev. Lett.} \textbf{\bibinfo{volume}{114}},
  \bibinfo{pages}{198301} (\bibinfo{year}{2015}{\natexlab{a}}).

\bibitem[{\citenamefont{Solon et~al.}(2015{\natexlab{b}})\citenamefont{Solon,
  Fily, Baskaran, Cates, Kafri, Kardar, and Tailleur}}]{solo15a}
\bibinfo{author}{\bibfnamefont{A.~P.} \bibnamefont{Solon}},
  \bibinfo{author}{\bibfnamefont{Y.}~\bibnamefont{Fily}},
  \bibinfo{author}{\bibfnamefont{A.}~\bibnamefont{Baskaran}},
  \bibinfo{author}{\bibfnamefont{M.~E.} \bibnamefont{Cates}},
  \bibinfo{author}{\bibfnamefont{Y.}~\bibnamefont{Kafri}},
  \bibinfo{author}{\bibfnamefont{M.}~\bibnamefont{Kardar}}, \bibnamefont{and}
  \bibinfo{author}{\bibfnamefont{J.}~\bibnamefont{Tailleur}},
  \bibinfo{journal}{Nature Phys.} \textbf{\bibinfo{volume}{11}},
  \bibinfo{pages}{673} (\bibinfo{year}{2015}{\natexlab{b}}).

\bibitem[{\citenamefont{Winkler et~al.}(2015)\citenamefont{Winkler, Wysocki,
  and Gompper}}]{wink15}
\bibinfo{author}{\bibfnamefont{R.~G.} \bibnamefont{Winkler}},
  \bibinfo{author}{\bibfnamefont{A.}~\bibnamefont{Wysocki}}, \bibnamefont{and}
  \bibinfo{author}{\bibfnamefont{G.}~\bibnamefont{Gompper}},
  \bibinfo{journal}{Soft Matter} \textbf{\bibinfo{volume}{11}},
  \bibinfo{pages}{6680} (\bibinfo{year}{2015}).

\bibitem[{\citenamefont{Bialk\'e
  et~al.}(2015{\natexlab{a}})\citenamefont{Bialk\'e, Siebert, L\"owen, and
  Speck}}]{bial15}
\bibinfo{author}{\bibfnamefont{J.}~\bibnamefont{Bialk\'e}},
  \bibinfo{author}{\bibfnamefont{J.~T.} \bibnamefont{Siebert}},
  \bibinfo{author}{\bibfnamefont{H.}~\bibnamefont{L\"owen}}, \bibnamefont{and}
  \bibinfo{author}{\bibfnamefont{T.}~\bibnamefont{Speck}},
  \bibinfo{journal}{Phys. Rev. Lett.} \textbf{\bibinfo{volume}{115}},
  \bibinfo{pages}{098301} (\bibinfo{year}{2015}{\natexlab{a}}).

\bibitem[{\citenamefont{Takatori and Brady}(2015{\natexlab{a}})}]{taka15}
\bibinfo{author}{\bibfnamefont{S.~C.} \bibnamefont{Takatori}} \bibnamefont{and}
  \bibinfo{author}{\bibfnamefont{J.~F.} \bibnamefont{Brady}},
  \bibinfo{journal}{Phys. Rev. E} \textbf{\bibinfo{volume}{91}},
  \bibinfo{pages}{032117} (\bibinfo{year}{2015}{\natexlab{a}}).

\bibitem[{\citenamefont{Takatori and Brady}(2015{\natexlab{b}})}]{taka15a}
\bibinfo{author}{\bibfnamefont{S.~C.} \bibnamefont{Takatori}} \bibnamefont{and}
  \bibinfo{author}{\bibfnamefont{J.~F.} \bibnamefont{Brady}},
  \bibinfo{journal}{Soft Matter} \textbf{\bibinfo{volume}{11}},
  \bibinfo{pages}{7920} (\bibinfo{year}{2015}{\natexlab{b}}).

\bibitem[{\citenamefont{Marconi and Maggi}(2015)}]{marc15}
\bibinfo{author}{\bibfnamefont{U.~M.~B.} \bibnamefont{Marconi}}
  \bibnamefont{and} \bibinfo{author}{\bibfnamefont{C.}~\bibnamefont{Maggi}},
  \bibinfo{journal}{Soft Matter} \textbf{\bibinfo{volume}{11}},
  \bibinfo{pages}{8768} (\bibinfo{year}{2015}).

\bibitem[{\citenamefont{Fily and Marchetti}(2012)}]{yaou12}
\bibinfo{author}{\bibfnamefont{Y.}~\bibnamefont{Fily}} \bibnamefont{and}
  \bibinfo{author}{\bibfnamefont{M.~C.} \bibnamefont{Marchetti}},
  \bibinfo{journal}{Phys. Rev. Lett.} \textbf{\bibinfo{volume}{108}},
  \bibinfo{pages}{235702} (\bibinfo{year}{2012}).

\bibitem[{\citenamefont{Palacci et~al.}(2013)\citenamefont{Palacci, Sacanna,
  Steinberg, Pine, and Chaikin}}]{pala13}
\bibinfo{author}{\bibfnamefont{J.}~\bibnamefont{Palacci}},
  \bibinfo{author}{\bibfnamefont{S.}~\bibnamefont{Sacanna}},
  \bibinfo{author}{\bibfnamefont{A.~P.} \bibnamefont{Steinberg}},
  \bibinfo{author}{\bibfnamefont{D.~J.} \bibnamefont{Pine}}, \bibnamefont{and}
  \bibinfo{author}{\bibfnamefont{P.~M.} \bibnamefont{Chaikin}},
  \bibinfo{journal}{Science} \textbf{\bibinfo{volume}{339}},
  \bibinfo{pages}{936} (\bibinfo{year}{2013}).

\bibitem[{\citenamefont{Buttinoni et~al.}(2013)\citenamefont{Buttinoni,
  Bialk\'e, K\"ummel, L\"owen, Bechinger, and Speck}}]{butt13}
\bibinfo{author}{\bibfnamefont{I.}~\bibnamefont{Buttinoni}},
  \bibinfo{author}{\bibfnamefont{J.}~\bibnamefont{Bialk\'e}},
  \bibinfo{author}{\bibfnamefont{F.}~\bibnamefont{K\"ummel}},
  \bibinfo{author}{\bibfnamefont{H.}~\bibnamefont{L\"owen}},
  \bibinfo{author}{\bibfnamefont{C.}~\bibnamefont{Bechinger}},
  \bibnamefont{and} \bibinfo{author}{\bibfnamefont{T.}~\bibnamefont{Speck}},
  \bibinfo{journal}{Phys. Rev. Lett.} \textbf{\bibinfo{volume}{110}},
  \bibinfo{pages}{238301} (\bibinfo{year}{2013}).

\bibitem[{\citenamefont{Stenhammar et~al.}(2013)\citenamefont{Stenhammar,
  Tiribocchi, Allen, Marenduzzo, and Cates}}]{sten13}
\bibinfo{author}{\bibfnamefont{J.}~\bibnamefont{Stenhammar}},
  \bibinfo{author}{\bibfnamefont{A.}~\bibnamefont{Tiribocchi}},
  \bibinfo{author}{\bibfnamefont{R.~J.} \bibnamefont{Allen}},
  \bibinfo{author}{\bibfnamefont{D.}~\bibnamefont{Marenduzzo}},
  \bibnamefont{and} \bibinfo{author}{\bibfnamefont{M.~E.} \bibnamefont{Cates}},
  \bibinfo{journal}{Phys. Rev. Lett.} \textbf{\bibinfo{volume}{111}},
  \bibinfo{pages}{145702} (\bibinfo{year}{2013}).

\bibitem[{\citenamefont{Stenhammar et~al.}(2014)\citenamefont{Stenhammar,
  Marenduzzo, Allen, and Cates}}]{sten14}
\bibinfo{author}{\bibfnamefont{J.}~\bibnamefont{Stenhammar}},
  \bibinfo{author}{\bibfnamefont{D.}~\bibnamefont{Marenduzzo}},
  \bibinfo{author}{\bibfnamefont{R.~J.} \bibnamefont{Allen}}, \bibnamefont{and}
  \bibinfo{author}{\bibfnamefont{M.~E.} \bibnamefont{Cates}},
  \bibinfo{journal}{Soft Matter} \textbf{\bibinfo{volume}{10}},
  \bibinfo{pages}{1489} (\bibinfo{year}{2014}).

\bibitem[{\citenamefont{Bialk\'e
  et~al.}(2015{\natexlab{b}})\citenamefont{Bialk\'e, Speck, and
  L\"owen}}]{bial14}
\bibinfo{author}{\bibfnamefont{J.}~\bibnamefont{Bialk\'e}},
  \bibinfo{author}{\bibfnamefont{T.}~\bibnamefont{Speck}}, \bibnamefont{and}
  \bibinfo{author}{\bibfnamefont{H.}~\bibnamefont{L\"owen}},
  \bibinfo{journal}{J. Non-Cryst. Solids} \textbf{\bibinfo{volume}{407}},
  \bibinfo{pages}{367â} (\bibinfo{year}{2015}{\natexlab{b}}).

\bibitem[{\citenamefont{Cates and Tailleur}(2015)}]{cate15}
\bibinfo{author}{\bibfnamefont{M.~E.} \bibnamefont{Cates}} \bibnamefont{and}
  \bibinfo{author}{\bibfnamefont{J.}~\bibnamefont{Tailleur}},
  \bibinfo{journal}{Annu. Rev. Condens. Matter Phys.}
  \textbf{\bibinfo{volume}{6}}, \bibinfo{pages}{219} (\bibinfo{year}{2015}).

\bibitem[{\citenamefont{Speck et~al.}(2015)\citenamefont{Speck, Menzel,
  Bialk\'e, and L\"owen}}]{spec15}
\bibinfo{author}{\bibfnamefont{T.}~\bibnamefont{Speck}},
  \bibinfo{author}{\bibfnamefont{A.~M.} \bibnamefont{Menzel}},
  \bibinfo{author}{\bibfnamefont{J.}~\bibnamefont{Bialk\'e}}, \bibnamefont{and}
  \bibinfo{author}{\bibfnamefont{H.}~\bibnamefont{L\"owen}},
  \bibinfo{journal}{J. Chem. Phys.} \textbf{\bibinfo{volume}{142}},
  \bibinfo{eid}{224109} (\bibinfo{year}{2015}).

\bibitem[{\citenamefont{Lee}(2013)}]{lee13a}
\bibinfo{author}{\bibfnamefont{C.~F.} \bibnamefont{Lee}}, \bibinfo{journal}{New
  J. Phys.} \textbf{\bibinfo{volume}{15}}, \bibinfo{pages}{055007}
  (\bibinfo{year}{2013}).

\bibitem[{\citenamefont{Elgeti and Gompper}(2013)}]{elge13}
\bibinfo{author}{\bibfnamefont{J.}~\bibnamefont{Elgeti}} \bibnamefont{and}
  \bibinfo{author}{\bibfnamefont{G.}~\bibnamefont{Gompper}},
  \bibinfo{journal}{EPL} \textbf{\bibinfo{volume}{101}}, \bibinfo{pages}{48003}
  (\bibinfo{year}{2013}).

\bibitem[{\citenamefont{Fily et~al.}(2015)\citenamefont{Fily, Baskaran, and
  Hagan}}]{fily15}
\bibinfo{author}{\bibfnamefont{Y.}~\bibnamefont{Fily}},
  \bibinfo{author}{\bibfnamefont{A.}~\bibnamefont{Baskaran}}, \bibnamefont{and}
  \bibinfo{author}{\bibfnamefont{M.~F.} \bibnamefont{Hagan}},
  \bibinfo{journal}{Phys. Rev. E} \textbf{\bibinfo{volume}{91}},
  \bibinfo{pages}{012125} (\bibinfo{year}{2015}).

\bibitem[{\citenamefont{Fily et~al.}(2014)\citenamefont{Fily, Baskaran, and
  Hagan}}]{fily14a}
\bibinfo{author}{\bibfnamefont{Y.}~\bibnamefont{Fily}},
  \bibinfo{author}{\bibfnamefont{A.}~\bibnamefont{Baskaran}}, \bibnamefont{and}
  \bibinfo{author}{\bibfnamefont{M.~F.} \bibnamefont{Hagan}},
  \bibinfo{journal}{Soft Matter} \textbf{\bibinfo{volume}{10}},
  \bibinfo{pages}{5609} (\bibinfo{year}{2014}).

\bibitem[{\citenamefont{Mallory et~al.}(2014)\citenamefont{Mallory, \ifmmode
  \check{S}\else \v{S}\fi{}ari\ifmmode~\acute{c}\else \'{c}\fi{}, Valeriani,
  and Cacciuto}}]{mall14}
\bibinfo{author}{\bibfnamefont{S.~A.} \bibnamefont{Mallory}},
  \bibinfo{author}{\bibfnamefont{A.}~\bibnamefont{\ifmmode \check{S}\else
  \v{S}\fi{}ari\ifmmode~\acute{c}\else \'{c}\fi{}}},
  \bibinfo{author}{\bibfnamefont{C.}~\bibnamefont{Valeriani}},
  \bibnamefont{and} \bibinfo{author}{\bibfnamefont{A.}~\bibnamefont{Cacciuto}},
  \bibinfo{journal}{Phys. Rev. E} \textbf{\bibinfo{volume}{89}},
  \bibinfo{pages}{052303} (\bibinfo{year}{2014}).

\bibitem[{\citenamefont{Yang et~al.}(2014)\citenamefont{Yang, Manning, and
  Marchetti}}]{yang14}
\bibinfo{author}{\bibfnamefont{X.}~\bibnamefont{Yang}},
  \bibinfo{author}{\bibfnamefont{M.~L.} \bibnamefont{Manning}},
  \bibnamefont{and} \bibinfo{author}{\bibfnamefont{M.~C.}
  \bibnamefont{Marchetti}}, \bibinfo{journal}{Soft Matter}
  \textbf{\bibinfo{volume}{10}}, \bibinfo{pages}{6477} (\bibinfo{year}{2014}).

\bibitem[{\citenamefont{Spellings et~al.}(2015)\citenamefont{Spellings, Engel,
  Klotsa, Sabrina, Drews, Nguyen, Bishop, and Glotzer}}]{spell15}
\bibinfo{author}{\bibfnamefont{M.}~\bibnamefont{Spellings}},
  \bibinfo{author}{\bibfnamefont{M.}~\bibnamefont{Engel}},
  \bibinfo{author}{\bibfnamefont{D.}~\bibnamefont{Klotsa}},
  \bibinfo{author}{\bibfnamefont{S.}~\bibnamefont{Sabrina}},
  \bibinfo{author}{\bibfnamefont{A.~M.} \bibnamefont{Drews}},
  \bibinfo{author}{\bibfnamefont{N.~H.~P.} \bibnamefont{Nguyen}},
  \bibinfo{author}{\bibfnamefont{K.~J.~M.} \bibnamefont{Bishop}},
  \bibnamefont{and} \bibinfo{author}{\bibfnamefont{S.~C.}
  \bibnamefont{Glotzer}}, \bibinfo{journal}{Proc. Natl. Acad. Sci. U.S.A.}
  \textbf{\bibinfo{volume}{112}}, \bibinfo{pages}{E4642}
  (\bibinfo{year}{2015}).

\bibitem[{\citenamefont{Smallenburg and L\"owen}(2015)}]{smal15}
\bibinfo{author}{\bibfnamefont{F.}~\bibnamefont{Smallenburg}} \bibnamefont{and}
  \bibinfo{author}{\bibfnamefont{H.}~\bibnamefont{L\"owen}},
  \bibinfo{journal}{Phys. Rev. E} \textbf{\bibinfo{volume}{92}},
  \bibinfo{pages}{032304} (\bibinfo{year}{2015}).

\bibitem[{\citenamefont{Wysocki et~al.}(2015)\citenamefont{Wysocki, Elgeti, and
  Gompper}}]{wyso15}
\bibinfo{author}{\bibfnamefont{A.}~\bibnamefont{Wysocki}},
  \bibinfo{author}{\bibfnamefont{J.}~\bibnamefont{Elgeti}}, \bibnamefont{and}
  \bibinfo{author}{\bibfnamefont{G.}~\bibnamefont{Gompper}},
  \bibinfo{journal}{Phys. Rev. E} \textbf{\bibinfo{volume}{91}},
  \bibinfo{pages}{050302} (\bibinfo{year}{2015}).

\bibitem[{\citenamefont{Nikola et~al.}(2016)\citenamefont{Nikola, Solon, Kafri,
  Kardar, Tailleur, and Voituriez}}]{niko16}
\bibinfo{author}{\bibfnamefont{N.}~\bibnamefont{Nikola}},
  \bibinfo{author}{\bibfnamefont{A.~P.} \bibnamefont{Solon}},
  \bibinfo{author}{\bibfnamefont{Y.}~\bibnamefont{Kafri}},
  \bibinfo{author}{\bibfnamefont{M.}~\bibnamefont{Kardar}},
  \bibinfo{author}{\bibfnamefont{J.}~\bibnamefont{Tailleur}}, \bibnamefont{and}
  \bibinfo{author}{\bibfnamefont{R.}~\bibnamefont{Voituriez}},
  \emph{\bibinfo{title}{Active particles on curved surfaces: Equation of state,
  ratchets, and instabilities}} (\bibinfo{year}{2016}),
  \bibinfo{note}{arXiv:1512.05697}.

\bibitem[{\citenamefont{Palacci et~al.}(2010)\citenamefont{Palacci,
  Cottin-Bizonne, Ybert, and Bocquet}}]{pala10}
\bibinfo{author}{\bibfnamefont{J.}~\bibnamefont{Palacci}},
  \bibinfo{author}{\bibfnamefont{C.}~\bibnamefont{Cottin-Bizonne}},
  \bibinfo{author}{\bibfnamefont{C.}~\bibnamefont{Ybert}}, \bibnamefont{and}
  \bibinfo{author}{\bibfnamefont{L.}~\bibnamefont{Bocquet}},
  \bibinfo{journal}{Phys. Rev. Lett.} \textbf{\bibinfo{volume}{105}},
  \bibinfo{pages}{088304} (\bibinfo{year}{2010}).

\bibitem[{\citenamefont{Enculescu and Stark}(2011)}]{encu11}
\bibinfo{author}{\bibfnamefont{M.}~\bibnamefont{Enculescu}} \bibnamefont{and}
  \bibinfo{author}{\bibfnamefont{H.}~\bibnamefont{Stark}},
  \bibinfo{journal}{Phys. Rev. Lett.} \textbf{\bibinfo{volume}{107}},
  \bibinfo{pages}{058301} (\bibinfo{year}{2011}).

\bibitem[{\citenamefont{Ginot et~al.}(2015)\citenamefont{Ginot, Theurkauff,
  Levis, Ybert, Bocquet, Berthier, and Cottin-Bizonne}}]{gino15}
\bibinfo{author}{\bibfnamefont{F.}~\bibnamefont{Ginot}},
  \bibinfo{author}{\bibfnamefont{I.}~\bibnamefont{Theurkauff}},
  \bibinfo{author}{\bibfnamefont{D.}~\bibnamefont{Levis}},
  \bibinfo{author}{\bibfnamefont{C.}~\bibnamefont{Ybert}},
  \bibinfo{author}{\bibfnamefont{L.}~\bibnamefont{Bocquet}},
  \bibinfo{author}{\bibfnamefont{L.}~\bibnamefont{Berthier}}, \bibnamefont{and}
  \bibinfo{author}{\bibfnamefont{C.}~\bibnamefont{Cottin-Bizonne}},
  \bibinfo{journal}{Phys. Rev. X} \textbf{\bibinfo{volume}{5}},
  \bibinfo{pages}{011004} (\bibinfo{year}{2015}).

\bibitem[{\citenamefont{Falasco et~al.}(2015)\citenamefont{Falasco, Baldovin,
  Kroy, and Baiesi}}]{fala16}
\bibinfo{author}{\bibfnamefont{G.}~\bibnamefont{Falasco}},
  \bibinfo{author}{\bibfnamefont{F.}~\bibnamefont{Baldovin}},
  \bibinfo{author}{\bibfnamefont{K.}~\bibnamefont{Kroy}}, \bibnamefont{and}
  \bibinfo{author}{\bibfnamefont{M.}~\bibnamefont{Baiesi}}
  (\bibinfo{year}{2015}), \bibinfo{note}{arXiv:1512.01687}.

\bibitem[{\citenamefont{Metropolis et~al.}(1953)\citenamefont{Metropolis,
  Rosenbluth, Rosenbluth, Teller, and Teller}}]{metr53}
\bibinfo{author}{\bibfnamefont{N.}~\bibnamefont{Metropolis}},
  \bibinfo{author}{\bibfnamefont{A.~W.} \bibnamefont{Rosenbluth}},
  \bibinfo{author}{\bibfnamefont{M.~N.} \bibnamefont{Rosenbluth}},
  \bibinfo{author}{\bibfnamefont{A.~H.} \bibnamefont{Teller}},
  \bibnamefont{and} \bibinfo{author}{\bibfnamefont{E.}~\bibnamefont{Teller}},
  \bibinfo{journal}{The Journal of Chemical Physics}
  \textbf{\bibinfo{volume}{21}}, \bibinfo{pages}{1087} (\bibinfo{year}{1953}).

\bibitem[{\citenamefont{Joyeux and Bertin}(2016)}]{joye16}
\bibinfo{author}{\bibfnamefont{M.}~\bibnamefont{Joyeux}} \bibnamefont{and}
  \bibinfo{author}{\bibfnamefont{E.}~\bibnamefont{Bertin}},
  \bibinfo{journal}{Phys. Rev. E} \textbf{\bibinfo{volume}{93}},
  \bibinfo{pages}{032605} (\bibinfo{year}{2016}).

\bibitem[{\citenamefont{Hansen and McDonald}(2006)}]{hansen}
\bibinfo{author}{\bibfnamefont{J.}~\bibnamefont{Hansen}} \bibnamefont{and}
  \bibinfo{author}{\bibfnamefont{I.}~\bibnamefont{McDonald}},
  \emph{\bibinfo{title}{Theory of Simple Liquids}}
  (\bibinfo{publisher}{Academic Press}, \bibinfo{address}{Amsterdam},
  \bibinfo{year}{2006}), \bibinfo{edition}{3rd} ed.

\bibitem[{\citenamefont{Howse et~al.}(2007)\citenamefont{Howse, Jones, Ryan,
  Gough, Vafabakhsh, and Golestanian}}]{hows07}
\bibinfo{author}{\bibfnamefont{J.~R.} \bibnamefont{Howse}},
  \bibinfo{author}{\bibfnamefont{R.~A.~L.} \bibnamefont{Jones}},
  \bibinfo{author}{\bibfnamefont{A.~J.} \bibnamefont{Ryan}},
  \bibinfo{author}{\bibfnamefont{T.}~\bibnamefont{Gough}},
  \bibinfo{author}{\bibfnamefont{R.}~\bibnamefont{Vafabakhsh}},
  \bibnamefont{and}
  \bibinfo{author}{\bibfnamefont{R.}~\bibnamefont{Golestanian}},
  \bibinfo{journal}{Phys. Rev. Lett.} \textbf{\bibinfo{volume}{99}},
  \bibinfo{pages}{048102} (\bibinfo{year}{2007}).

\bibitem[{\citenamefont{Thompson and Jack}(2015)}]{thom15}
\bibinfo{author}{\bibfnamefont{I.~R.} \bibnamefont{Thompson}} \bibnamefont{and}
  \bibinfo{author}{\bibfnamefont{R.~L.} \bibnamefont{Jack}},
  \bibinfo{journal}{Phys. Rev. E} \textbf{\bibinfo{volume}{92}},
  \bibinfo{pages}{052115} (\bibinfo{year}{2015}).

\bibitem[{\citenamefont{Ezhilan et~al.}(2015)\citenamefont{Ezhilan,
  Alonso-Matilla, and Saintillan}}]{ezhi15}
\bibinfo{author}{\bibfnamefont{B.}~\bibnamefont{Ezhilan}},
  \bibinfo{author}{\bibfnamefont{R.}~\bibnamefont{Alonso-Matilla}},
  \bibnamefont{and}
  \bibinfo{author}{\bibfnamefont{D.}~\bibnamefont{Saintillan}},
  \bibinfo{journal}{J. Fluid Mech.} \textbf{\bibinfo{volume}{781}}
  (\bibinfo{year}{2015}).

\bibitem[{\citenamefont{Yan and Brady}(2015{\natexlab{b}})}]{yan15a}
\bibinfo{author}{\bibfnamefont{W.}~\bibnamefont{Yan}} \bibnamefont{and}
  \bibinfo{author}{\bibfnamefont{J.~F.} \bibnamefont{Brady}},
  \bibinfo{journal}{J. Fluid Mech.} \textbf{\bibinfo{volume}{785}},
  \bibinfo{pages}{R1} (\bibinfo{year}{2015}{\natexlab{b}}).

\bibitem[{\citenamefont{Harder et~al.}(2014)\citenamefont{Harder, Mallory,
  Tung, Valeriani, and Cacciuto}}]{hard14}
\bibinfo{author}{\bibfnamefont{J.}~\bibnamefont{Harder}},
  \bibinfo{author}{\bibfnamefont{S.~A.} \bibnamefont{Mallory}},
  \bibinfo{author}{\bibfnamefont{C.}~\bibnamefont{Tung}},
  \bibinfo{author}{\bibfnamefont{C.}~\bibnamefont{Valeriani}},
  \bibnamefont{and} \bibinfo{author}{\bibfnamefont{A.}~\bibnamefont{Cacciuto}},
  \bibinfo{journal}{J. Chem. Phys.} \textbf{\bibinfo{volume}{141}},
  \bibinfo{pages}{194901} (\bibinfo{year}{2014}).

\bibitem[{\citenamefont{Ray et~al.}(2014)\citenamefont{Ray, Reichhardt, and
  Reichhardt}}]{ray14}
\bibinfo{author}{\bibfnamefont{D.}~\bibnamefont{Ray}},
  \bibinfo{author}{\bibfnamefont{C.}~\bibnamefont{Reichhardt}},
  \bibnamefont{and} \bibinfo{author}{\bibfnamefont{C.~J.~O.}
  \bibnamefont{Reichhardt}}, \bibinfo{journal}{Phys. Rev. E}
  \textbf{\bibinfo{volume}{90}}, \bibinfo{pages}{013019}
  (\bibinfo{year}{2014}).

\bibitem[{\citenamefont{Ni et~al.}(2015)\citenamefont{Ni, Cohen~Stuart, and
  Bolhuis}}]{ran15}
\bibinfo{author}{\bibfnamefont{R.}~\bibnamefont{Ni}},
  \bibinfo{author}{\bibfnamefont{M.~A.} \bibnamefont{Cohen~Stuart}},
  \bibnamefont{and} \bibinfo{author}{\bibfnamefont{P.~G.}
  \bibnamefont{Bolhuis}}, \bibinfo{journal}{Phys. Rev. Lett.}
  \textbf{\bibinfo{volume}{114}}, \bibinfo{pages}{018302}
  (\bibinfo{year}{2015}).

\bibitem[{\citenamefont{Farage et~al.}(2015)\citenamefont{Farage, Krinninger,
  and Brader}}]{fara15}
\bibinfo{author}{\bibfnamefont{T.~F.~F.} \bibnamefont{Farage}},
  \bibinfo{author}{\bibfnamefont{P.}~\bibnamefont{Krinninger}},
  \bibnamefont{and} \bibinfo{author}{\bibfnamefont{J.~M.}
  \bibnamefont{Brader}}, \bibinfo{journal}{Phys. Rev. E}
  \textbf{\bibinfo{volume}{91}}, \bibinfo{pages}{042310}
  (\bibinfo{year}{2015}).

\bibitem[{\citenamefont{Maggi et~al.}(2015)\citenamefont{Maggi, Marconi, Gnan,
  and Di~Leonardo}}]{magg15}
\bibinfo{author}{\bibfnamefont{C.}~\bibnamefont{Maggi}},
  \bibinfo{author}{\bibfnamefont{U.~M.~B.} \bibnamefont{Marconi}},
  \bibinfo{author}{\bibfnamefont{N.}~\bibnamefont{Gnan}}, \bibnamefont{and}
  \bibinfo{author}{\bibfnamefont{R.}~\bibnamefont{Di~Leonardo}},
  \bibinfo{journal}{Sci. Rep.} \textbf{\bibinfo{volume}{5}},
  \bibinfo{pages}{10742} (\bibinfo{year}{2015}).

\bibitem[{\citenamefont{Marchetti et~al.}(2016)\citenamefont{Marchetti, Fily,
  Henkes, Patch, and Yllanes}}]{marc16}
\bibinfo{author}{\bibfnamefont{M.~C.} \bibnamefont{Marchetti}},
  \bibinfo{author}{\bibfnamefont{Y.}~\bibnamefont{Fily}},
  \bibinfo{author}{\bibfnamefont{S.}~\bibnamefont{Henkes}},
  \bibinfo{author}{\bibfnamefont{A.}~\bibnamefont{Patch}}, \bibnamefont{and}
  \bibinfo{author}{\bibfnamefont{D.}~\bibnamefont{Yllanes}},
  \bibinfo{journal}{Curr. Opin. Colloid Interface Sci.} pp.~\bibinfo{pages}{--}
  (\bibinfo{year}{2016}).

\bibitem[{\citenamefont{Wittkowski et~al.}(2014)\citenamefont{Wittkowski,
  Tiribocchi, Stenhammar, Allen, Marenduzzo, and Cates}}]{witt14}
\bibinfo{author}{\bibfnamefont{R.}~\bibnamefont{Wittkowski}},
  \bibinfo{author}{\bibfnamefont{A.}~\bibnamefont{Tiribocchi}},
  \bibinfo{author}{\bibfnamefont{J.}~\bibnamefont{Stenhammar}},
  \bibinfo{author}{\bibfnamefont{R.~J.} \bibnamefont{Allen}},
  \bibinfo{author}{\bibfnamefont{D.}~\bibnamefont{Marenduzzo}},
  \bibnamefont{and} \bibinfo{author}{\bibfnamefont{M.~E.} \bibnamefont{Cates}},
  \bibinfo{journal}{Nat. Commun.} \textbf{\bibinfo{volume}{5}},
  \bibinfo{pages}{4351} (\bibinfo{year}{2014}).

\bibitem[{\citenamefont{Tiribocchi et~al.}(2015)\citenamefont{Tiribocchi,
  Wittkowski, Marenduzzo, and Cates}}]{tiri15}
\bibinfo{author}{\bibfnamefont{A.}~\bibnamefont{Tiribocchi}},
  \bibinfo{author}{\bibfnamefont{R.}~\bibnamefont{Wittkowski}},
  \bibinfo{author}{\bibfnamefont{D.}~\bibnamefont{Marenduzzo}},
  \bibnamefont{and} \bibinfo{author}{\bibfnamefont{M.~E.} \bibnamefont{Cates}},
  \bibinfo{journal}{Phys. Rev. Lett.} \textbf{\bibinfo{volume}{115}},
  \bibinfo{pages}{188302} (\bibinfo{year}{2015}).

\end{thebibliography}
\end{document}